\documentclass{article}
\usepackage{graphicx}
\usepackage{amsmath}

\begin{document}
\pagestyle{plain}
\setcounter{page}{0}

\parindent 8mm
\pagestyle{empty}

\vspace*{2cm}
\title{The Analyses of Node Swapping Networks by New Graph Index }
\author{ Norihito Toyota}
\date{Hokkaido Information University,
59-2 Nishinopporo Ebetsu City, Japan \\
E-mail:toyota@do-johodai.ac.jp}

\maketitle
\baselineskip 5mm
\pagestyle{plain}
\begin{abstract}
We have proposed two new dynamic networks where 
two node are swapped each other, and showed that the both networks behave as a small world like 
in the average path length but can not make any effective discussions on the clustering coefficient 
because of the topological invariant properties of the networks. 
In this article we introduce a new index, "hamming coefficient" or "multiplicity", that can act well 
  for these dynamic networks. 
The hamming coefficient or multiplicity is shown essentially to behave as the clustering coefficient in 
the small world network proposed by Watts and Strogatz\cite{Watt1}. 
By evaluating the new index, we uncover another properties of the two networks.   
\end{abstract}
\bf{key words:} 
\it{Small world network, Scale free network, node swapping network, 
clustering coefficient, haming distance, multiplicity  }

\rm
\normalsize
\baselineskip 6mm
\pagestyle{plain}

\renewcommand{\thefootnote}{\fnsymbol{footnote}} 
\rm
\normalsize
\baselineskip 6mm

\section{Introduction}
In a social network, we may need to consider the possibility that people is transferred 
to another place. 
Then the physical (direct) relations among them are often lost by the movement. 
In terms of a network theory, this means that some nodes break the present connections 
with neighboring nodes, move and there build new connections with nodes.
  For simplicity,  we here consider only that two nodes exchange the place each other on the network. 
Such exchange is assumed to be constantly carried out.  
Some properties such as the diameter, the average path length, 
the propagation when one virus is placed on the network, 
have been studied by the author\cite{toyota1} 
where it has been pointed out that the swapping networks look like a little small world property 
but have rather intermedidate properties 
between the small world network (SW-NET) introduced 
by Watts and Strogatz\cite{Watt1},\cite{Watt2} 
and regular lattices.  
In this article, we study the dynamic networks in more details. 

The clustering coefficient is used in usual network analyses. 
There are, however,  two difficult points in estimating it  
in node swapping networks (NSN). 
First is that general dynamic networks such as swapping networks has 
a time dependent clustering coefficient, unlike ststic networks such as SW-NET or 
the preferencial scale free model (SF-NET) introduced 
by Barabasi and Albert\cite{Albe1},\cite{Albe2},\cite{Albe3}, 
where they are usually alalyzed after networks are completed.   
 Second is that the topology of the NSN is invariant under the swapping of nodes,  
because swapped nodes inherit the all links connected with old nodes after swapping,  
and so  the clustering coefficient is trivially constant. 
Thus we need to introduce a sort of new index corresponding to the clustering coeffient. 
In any dynamic networks, it would be crucial that some propagation process is considered. 
We consider the situation that  a test virus is randomly placed on a node on  NSN.   
 The virus propagates a next one connected with the first target node. 
We compare the similarity of the frendships of two nodes, the first node $i$ and propagated 
node $j$, that is to say, estimate the hamming distance between the two nodes 
with adjacent vectors, $v_i$ and $v_j$ that are the $i-$th row component and 
$j-$th row component of the adjacent matrix corresponding to the network, respectively. 
The estimated quantity is related only to the simirality of two connected nodes. 
The clustering coefficient are related to the similarity of all nodes connected with a target node. 
So the new index can interpreted to be  a sort of the shortening of the clustering coefficient.   
A virus propagates from the target node to connected nodes one after another and 
we estimate the index every time the virus infects some connected node.  
We consider the average hamming distance of them duaring propagation 
till all nodes are infected. 
We compare the average hamming distance with  the clustering coefficient of SW-NET.   
As result we turn out that they show the similar behaviours as the rewiring probability increases.   
So this suggests that we can use this new index instead of 
the clustering coefficient in dynamic networks. 
By evaluating the new index,   
we can observe that the NSN certainly behaves as SW-NET in the average 
hamming distance, unlike in an average path length. 
Thus we conclude that the NSN is not so small world like as SW-NET but looks like SW-NET 
with respect to this new index corresponding to the usual clustering coefficient.

This article is planed as follows. 
First section is devoted to the Introduction, and we give a brief review 
of the swapping network togrther with the constructive definition 
 that was given in the previous article\cite{toyota1}. 
In section 3 we give the definition of the new index, average hamming distance and multiplicity,
and compare it with the clustering coefficient of the well known SW-NET.  
After that, we evaluate the indeces for NSN and the preferential NSN \cite{toyota1} 
to study the properties of the network in more details in the section 4.  
In the last section, 5, we give concluding remarks. 

\section{Node swapping netwark and average path length between nodes}
In this section we introduce the NSN by presenting a constructive definition,  
and review some properties of the network that have been discussen in \cite{toyota1}. 

\subsection{Review of node swapping network (NSN)}
As explained in the previous section, we consider that nodes swap each other on a regular network. 
This network may be  seemed to look like a small world network. 
However, it is necessarily not the case. 
By the movement, 
the nodes and the edges accompanied with them are entirely cut, 
and the nodes are connected with new edges each other at the new position.   
Notice that the network topology is apparently invariant under the procedure. 
In  small world networks the  static properties are only pursured but  
the  dynamic properties such as NSN are rather important.   

The algorithm for formulating the NSN is as follows;\\
1. Prepare a regular (typically one dimensional) network with a periodic boundary condition 
such as a  ring. \\
2. Randomly choose two nodes on its network and swap them. 
This procedure is repeated $Q$ times.\\
3. Evaluate correct quantities of the network. \\
4. 1$\sim$3, which is one round,  is repeated $M$ times.   \\  
  
In such a way, the network is dynamically analyzed as edges are cut and pasetd to new nodes. 

We analyze some properties of the network by doing computer simulations. 
First of all we discuss  the diameter $D$ and 
the average distance $L$ between any pairs of nodes of the NSN,  
which have been given in \cite{toyota1}. 

The diameter of usual random networks behaves as $\frac{\log n}{\log <k>}$\cite{Chun},  
where $<k>$ is the average degree of nodes 
and $n$ is the size of network, that is, the number of nodes.    
We have conveniently introduced a handy network in \cite{toyota1} with the same properties essentially 
as random networks, instead of usual random networks. 
This new network has been called "random graph with fixed degree", RNFD,  
where the degrees of all nodes are contrived to be a constant number $k$.

 Fig.1 shows the size $n$ vs. $D$  of the NSN constructed from 
degree $k=4$ regular lattice  and  RNFD with $k=4$, respectively\cite{toyota1}.   
The points and curved lines in the figures show simulation data and its approximate curves, respectively.    
This shows that $n$ dependence of $D$ in the NSN is exponential, 
while that in RAFD is logarithmic such as random networks. 
Their essential properties are independent of $Q$ in NSN or $k$ in RNFD.  
Since the behaviour is linear, $D=\frac{n}{2k}$, in one dimensional regural lattice 
with the periodic boundary condition, it turns out that NSN  is a network 
intermediate between  regular lattices and random networks or SW-NET. 
The existence of $D$ also means that NSN is an overall connected network.  
\addtocounter{figure}{0}
\begin{figure}[b]
\begin{center}
\includegraphics[width=12cm,clip]{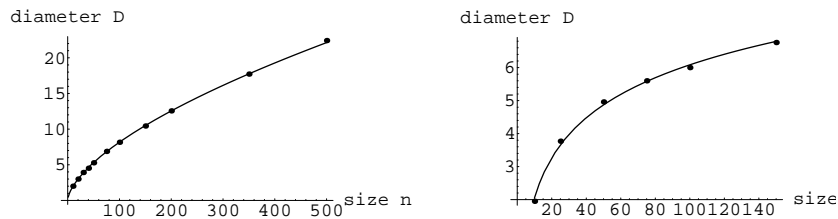}
\caption{Diameters of the NEN with $Q=10$ for average of $50$ times (left) and RNFD with $k= 4$ 
for average of $100$ times (right). Approximate formula of them are $D=0.4725 n^{0.619}$ 
and $D=1.7507 \log _e n -1.9778$, respectively. }
\end{center}
\end{figure}

To  clear the point we study average distance $L$ between any pairs of nodes.  
(Notice that the behaviour of $L$ is not necessarily equal to that of $D$ in dynamic networks, 
because the network in calculating the distances from a target node to nearby nodes 
is not the same as that in calculating the distance from the target node to faraway one.  
Thus $L$ is the average over  different networks. 
$D$ is the step number from a target one to the most faraway node.) 
Fig.2 shows $L$-$n$ curves of RNFD with $k=4$ and NSN with $Q=10$ and $k=4$.  
Essentially $L$'s have the same property as $D$. 
The reason will be that the number of steps needed for the complete estimation of $D$ 
is nearly equal to that of $L$.  
The properties are also independent of $Q$ or $k$. 
Since a regular lattice shows linear dependence in $L$-$n$ relation such as $D$-$n$,  
the  NSN is not only so small world and but also so large world after all.  
Fig. 3 refers to theoretical $L$-$n$ curves of SW-NET\cite{Newm1} and 
SF-NET\cite{Boll},\cite{Cohe} that are given by 
\begin{equation}
L(n)=\left\{
\begin{array}{ll}
\frac{\log(4np)}{8p} & \mbox{ for $2np >> 1$ and SW-NET}, \\
\frac{\log_e n}{ \log_e \log_e n} & \mbox{ for SF-NET},
\end{array}
\right.
\end{equation}
and their numerically approximated curves. 
$p$ is the rewiring probability in SW-NET, taken $p=0.05$ in Fig.3.   
In SF-NET, the logarithmic function phenomenologically fits almost perfectly. 
Though it is also  possible that both of NSN and SF-NET can be approximated by exponential functions, 
they are  very different from each other in the absolute value of the index.    
This property is essentially invariant under changing $Q$ value. 
As we increse $Q=1,\;5,\;  10, ...$, the index of the exponential decreses to 
$s=\;0.83,\; 0.62,\;0.58,...$ in NSN. 
As for SF-NET $s=0.07$, different from those of NEN in order,  
and it  seems not to be able to overcome  the difference 
(we should interpret that the excessively small $s$ means that it is rather the logarithmic function). 

Thus SF-NET and NSN are essentially thought to be different networks in terms of 
the average path length. 
In summary we conclude that the relation 
$$RNFD \sim SF-NET < SW-NET < NSN < Redular \;\;lattice$$
applys in $L$.  
\addtocounter{figure}{0}
\begin{figure}[t]
\begin{center}
\includegraphics[scale=1.0]{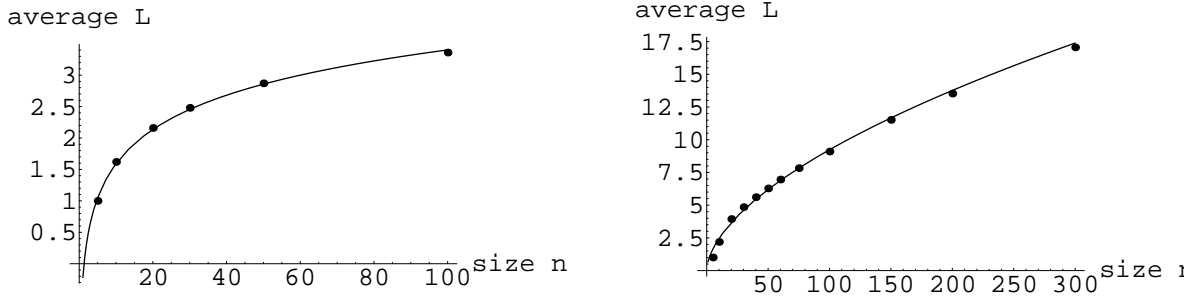}
\caption{Average distances between two nodes for average of $100$ times: 
The left is an average $L$ of RNFD with $k=4$. The right is that of NEN with $Q=10$ and $k=4$. 
The approximate formula of them are $L=0.7861 \times \log_e n- 0.2182$ and $L= 0.6509 \times n^{0.579}$, respectively}
\end{center}
\end{figure}

\addtocounter{figure}{0}
\begin{figure}[t]
\begin{center}
\includegraphics[scale=1.0]{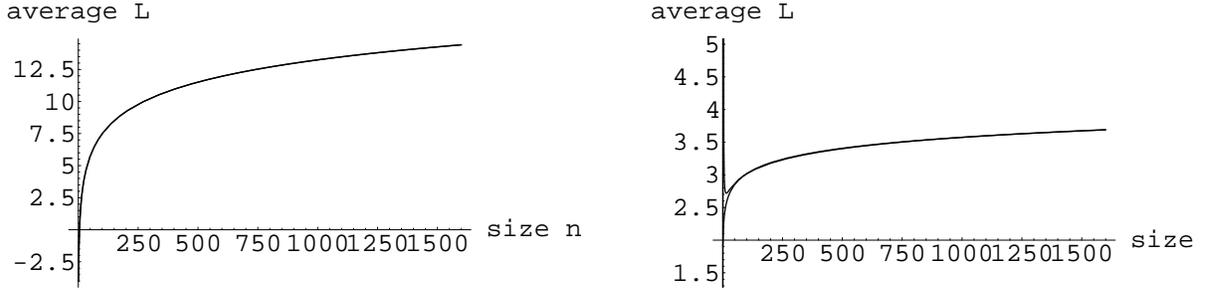}
\caption{Average distances between two nodes: The left is an average $L$ of SW-NET with $k=4$ and $p=0.05$, 
and the right is  that of SF-NET.  The approximate formula of them are $L=2.5 \log_e n- 4.0236$ 
and $L= 0.2421 \log_e n + 1.9031$ or $L=2.1888 \times n^{0.0707}$, respectively}
\end{center}
\end{figure}

\section{Hamming coefficient and clustering coefficient}
The clustering coeffient and the degree distribution  have no significance 
in the NSN, because the network topology in NSN is apparently invariant temporally 
so that they take the same values as those of the original regular lattice. 
As for this, we may have to introduce a sort of new kind of index to investigate 
NSN in more details. 

The most effective way would be to explore the propagation of a test virus on  dinamic networks. 
We adopt the idea, basically.  
Instead of exploring the similarities of friendship among all nodes connected with a target node 
such as the clustering coefficient,   
we estimate the similarity of friendship between a node $i$
connected with a target node $j$ and  the target node.  
  We measure it by calculating the hamming distance between adjacent vectors $v_i$ and $v_j$
 where the adjacent vector $v_i$($v_j$) is the $i$-th($j$-th) row vector 
in the adjacent matrix of the network.  
Then node $i$ is chosen at random among the conected nodes with $j$, which reflects the situation  
that a virus randomly infects some node connected with the target node $j$. 
 In place of the usual clustering coefficient, we evaluate the averaged hamming distance $D_H$
 during the time all nodes will be infected.  
Moreore exactly, we introduce the multiplicity $M$ as
\begin{equation}
M=1-\frac{D_H}{D_n},\;\;\;\; D_H=v_i \bullet v_j\;\;\mbox{ (where $ \bullet$ means the Boolean inner product)}
\end{equation} 
in order to measure a similarity of two nodes, 
while the hamming distance itself means the difference of friendship between two connected nodes. 
We take $D_n = 2k$ as the normarization factor (The reason will be given later). 

Next we compare the multiplicity to the usual clustering coefficient in well known 
networks such as the SW-NET.  
In Fig. 4  the two indeces in the SW-NET with degree $k=12$ and $n=500$   are given. 
The fact that both act in a similar way suggests that the multiplicity can play the same role 
as the clustering coefficient.  
Of course both indeces are originally defferent ones and so it is not necessary that they 
take a same value or behave in same way exactly. 
The multiplicity is only a substitute for the clustering coefficient. 
However, it can play an important role in dynamic networks sucu as NSN 
as discussed in the next section.

\addtocounter{figure}{0}
\begin{figure}[t]
\begin{center}
\includegraphics[scale=1.0]{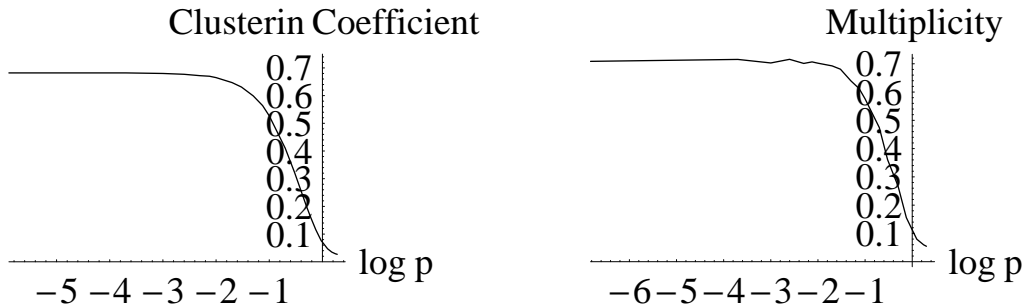}
\caption{Clustering Coefficient and Multiplicity of the SW-network with $\delta=6$ and $n=500$ 
averaged for 500 times.}
\end{center}
\end{figure}


Here we have a little theoretical discussion on $D_H$ to speculate the value of $D_n$. 
In regular lattice, we can analytically estimate $D_H$;  
\begin{equation}
D_H=\sum_{i=1}^{\delta} \frac{2i}{\delta}=\delta +1
\end{equation} 
where $k=2\delta$. 
This corresponds to the limit of $p \; \rightarrow \;0$ in SW-NET. 
The fact that $D_H=\delta+1=7$ for $\delta =6$ agrees with Fig.5 where $D_H=7.16$. 

On the other hand, in random lattice, we can estimate $D_H$ as an expectation value of  
the probability that $i$-th element in an $n$ bit string, whose component randomly takes 0 or 1,   
 is different from $i$-th one in another random $n$ bit string. 
So we obtain  
\begin{equation}
D_H=[ 1-\{ (\frac{2\delta}{n})^2 + (\frac{n-2\delta}{n})^2 \} ] \times n 
= 4\delta \frac{(n-\delta)}{n}
\end{equation} 
where the inner parts of $\{  \;\;  \}$ is the sum of two probabilities 
 that both $i$-th elements are 0 and that they are 1 together. 
Lastly $n$ is multiplied to take an average for $n$ bits.   
 Simply we can also evaluate it as the expectation value of the probability  
that $i$-th elements of two random $n$-bits strings  
are different each other;
 \begin{equation}
D_H= 2 \times n \times \frac{n-2\delta}{n} \frac{2\delta}{n}
= 4\delta \frac{(n-\delta)}{n}
\end{equation} 
where $n$ is multiplied to  take the average for $n$ bits as before and 
2 is multiplied due to the permutaion symmetry of two $n$-bit strings. 
More elaborate derivation will be given in Appendix. 
This happens at large $p$ for SW-NET and so $D_H =23.8 $ in the present case 
with $\delta =6$ and $n=500$, which agrees well with Fig.5 where $D_H=22.7$.

Anyway $\frac{4\delta (n-\delta)}{n}$ is the maximal value of $D_H$.  
For $n>>\delta$, $\frac{4\delta (n-\delta)}{n} \sim 4\delta$. 
Thus the normalization factor $D_n =2k=4\delta$ is taken in the equation (2). 
This ensures $0 \leq M \leq 1$.


\addtocounter{figure}{0}
\begin{figure}[h]
\begin{center}
\includegraphics[scale=1.0]{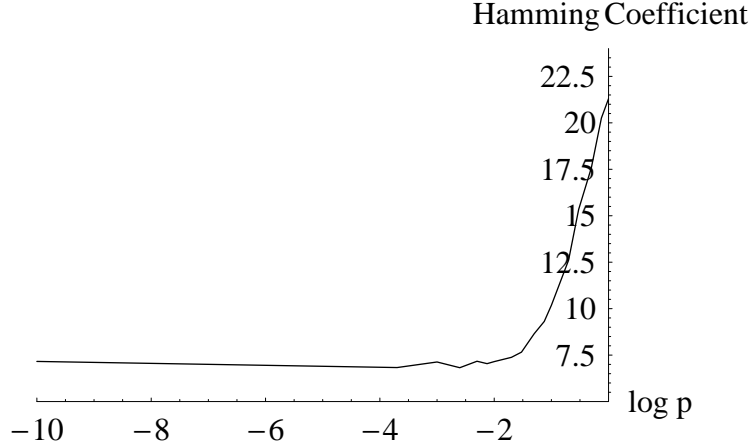}
\caption{Hamming coefficient of SW-NET with $\delta=6$ averaged for 100 times}
\end{center}
\end{figure}

\section{Multiplicity of simple node swapping network and 
prefferencial node swapping network}
In this section, we estimate the multiplicities of NSN and their variation, which will be 
defined in 4.2, to analyse network properties in more details.

\subsection{Hamming coefficient of simple node swapping network }
In this section we evaluate new index, the hamming distance, for NSN. 
For it,  we need a little extension of the index so that 
it adapts in dynamic networks.   
In dynamic networks, we define the incoming edges as the edges that a target node $i$ leaves,   
and the outgoing edges as those that the node $h$ infected from the node $i$ leaves. 
When no swapping happens, the index can be obtained by calculating  $v_i \bullet v_h$.   
However, notice that when the infected node $h$ is swapped, outgoing edges are different from 
those without swapping beacause of rewiring effect. 
When the infected node $h$ is swapped, the outgoing edges are those that 
the infected node gets at the new position.  
Then the hamming distance turns to the Boolean inner product betweem the target node $i$ and 
the node $j$ that is swapped with  the node $h$  connected with $i$, that is   
$v_i \bullet v_j$, ultimately. 
Thus we evaluate it for NSN and the preferential NSN, 
which will be explained in the  successive subsection.

\subsection{Multiplicities of simple and preferrencial node swapping network }
First of all we explain a variation of the NSN. 
There are a little similarity between the NSN and SF-NET apparently as suggested before. 
We pursue this point still more.  
Scale free property usually appears from both of the evolution and the preferential attachment.  
We apply  the idea  of the preferential attachment to this dynamic NSN. 
We assume that the nodes which has been transferred once are also transferred 
with high propability after that. 
At $m$ round and q times,  the probability  $p_i(t)$  that a node $i$ 
is chosen as a swapping node is assumed that
\begin{equation}
p_i(t)=\left\{
\begin{array}{ll}
\frac{1+p_i(t-1)N(t-1)}{N(t-1)+2}& \mbox{when the node $i$ was chosen as exchange node at time $t-1$ }, \\
p_i(t-1) & \mbox{others},
\end{array}
\right.
\end{equation}
where
\begin{equation}
N(t)=n+2t,\;\;\;\;t=m Q+q, \;\;\;\; 
p_i(0) =\frac{1}{N(0)}=\frac{1}{n} \; \mbox{ for all $i$}. 
\end{equation}
This reflects the fact that while active people often transfer, others trend to stay in one place.        
We call this type of networks \textit{Preferential Node Swapping Network} (PNSN). 
On the other hand, NSN introduced in the previous subsection is called \textit{simple NSN} 
when we need to distinguish them. 
The results of computer simulation of $L$ and $D$ on PNSN are just similar to those of the NEN
\cite{toyota1}.

Fig. 6 shows the multiplicity of the simple NSN and PNSN with $k=4$, $n=500$ and $Q=5$. 
More simulations will prove that changing $Q$ does not have any crucial effects in the multiplicity.   
We can observe that the behaviour of NSN is the almost  same as that of simple NSN.  
So it seems that there is not preferential effect in NSN,  
even when swapping increases in a number of times, which corresponds to large $p$.

From Fig.4 and Fig.6 where the multiplicity showly drops off in the similar manner as SW-NET, 
we can observe that the behaviour of (P)NSN looks like that of SW-NET in $M$. 
This means that NSN is definitely different  from SF-NET. 

\addtocounter{figure}{0}
\begin{figure}[h]
\begin{center}
\includegraphics[scale=1.0]{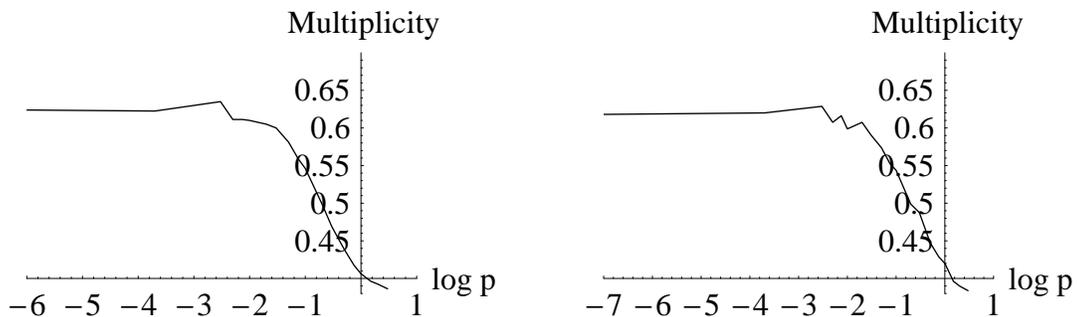}
\caption{Multiplicity of the simple NSN and PNSN with $k=4$, $n=500$ and $Q=5$. }
\end{center}
\end{figure}

\section{Concluding Remarks}

We introduced a new index for dynamic networks to analyze them, especially NSN or PSNS, in more details. 
It has been shown that this index, multiplicity, can stand in for the usual clustering coefficient
 in the SW-NET.  
Using this fact,  we analyse the simple NSN and the PNSN. 
These behaviours look like SW-NET in the point of view of the new index, $M$.  
Considering the results of analyses of the diameters and the average path length given in \cite{toyota1}, 
  we entirely obtain three main conclusions. 
One is that  NSN is not so small world as SW-NET and SF-NET, but 
a little more small world than regular lattice networks, and 
  thus NSN is something between regural networks and SW-NET.  
Second one is that (P)NSN shows similar behabiours as SW-NET in the multiplicity. 
Third is that there is not any crucial preferential effects in NSN. 

They are summarized with other well-known networks in Table 1 where the multiplicity in exchange for 
 $C$ is shown for (P)NSN.  
The properties of $L$ and $C$ of all networks that have already known currently \cite{masuda}
are included in the Table 1 except for (P)NSN.   
For example, the properties of complete graphs are essentially the same as those of SW-NET 
with respect to $L$ and $C$, and so on. 
By contrast, (P)NSN are different from every one of them that have already known. 
Moreover as $p \rightarrow$ large, $L$ increases and the $M$ corresponding to $C$ decreases in (P)NSN. 
By taking large $p$,  a network with large $L$ and small $M$($C$) may be constructed, which 
 has quire novel property.   
To study some dynamics of NSN with these properties will be next
intersting works\cite{matsuzaki1}.\\

\begin{table}[t]
\caption{Comparison of various networks with (P)NSN. }
\begin{center}
\begin{tabular}{|l|c|c|c|c|c|}\hline
  & Randm Networks & SF-NET & SW-NET ($0<p<1$) & (P)NSN  & Regular Lattice \\ \hline
L & small          & small &   small           &  middle & large \\ \hline
C(M) & small          & small &   large           &  large  & large \\ \hline
\end{tabular}
\end{center} 
\end{table} 

\subsection*{Acknowledgment}
I tahnk for R. Abe and U. Matsuzaki, especially S. Hayakawa for useful discussions. 

\appendix
\section*{Appendix }
\subsection*{Analitic derivation of hamming coefficient in  a random lattice}
We consider a network that the number of nodes is $n$  and  the degree of  nodes is $2 \delta =k$.   
The hamming coefficient $D_H$ in  a random lattice can be derived as follows; 
\begin{equation}
D_H= \sum_{m=0}^{n-2\delta} \frac{2m  \binom{2\delta}{m}\times \binom{n-2\delta}{m}}{A},
\end{equation} 
where $\binom{n}{m}$ shows the combination 
\begin{equation}
\binom{n}{m} =\frac{n!}{m!(n-m)!}, 
\end{equation}
and $A$ is a normalization factor;
\begin{equation}
A= \sum_{m=0}^{n-2\delta}  \binom{2\delta}{m} \times \binom{n-2\delta}{m}
  = \binom{n}{n-2\delta}  = \binom{n}{2\delta}. 
\end{equation}

Then we lead to the following equation that is exactly  same as equation (4) or (5); 

\begin{eqnarray}
D_H&=& \frac{1}{A} \sum_{m=0}^{n-2\delta} 2m  \binom{2\delta}{m}\times 
   \frac{(n-2\delta)!}{m!(n-2\delta -m)!}\\
 &=&\frac{1}{A} \sum_{m=0}^{n-2\delta} 2  \binom{2\delta}{m}\times 
   (n-2\delta)\frac{(n-2\delta-1)!}{(m-1)!(n-2\delta -m)!}\\
&=&\frac{1}{A} \sum_{m=0}^{n-2\delta} 
        2 \binom{2\delta}{m}\times \binom{n-2\delta-1}{m-1}\\
&=&\frac{2(n-2\delta)}{A} \binom{n-1}{n-2\delta}\\
&=& 2(n-2\delta) \frac{(n-2\delta)!(2\delta)!}{n!} \frac{(n-1)!}{(n-2\delta)!(2\delta-1)!}\\
&=&4\delta \frac{(n-2\delta)}{n},
\end{eqnarray} 

where we used the  following  formulus;

 \begin{equation}
 \sum_{m=0}^{q}  \binom{X}{m} \times \binom{Y}{p-m}
  = \binom{X+Y}{q}. 
\end{equation}




\end{document}